\documentclass[12pt]{article}
\input epsf

\def\plottwo#1#2{\centering \leavevmode
\epsfxsize= #2\columnwidth \epsfbox{#1}}

\def\apj{Astrophys. J.~}
\def\apjl{Astrophys. J. Lett. }
\def\prd{Phys. Rev. D }
\def\prl{Phys. Rev. Lett. }
\def\mnras{Mon. Not. R. astron. Soc. }

\def\aj{Astron. J. }

\def\be{\begin{equation}}
\def\ee{\end{equation}}
\def\bea{\begin{eqnarray}}
\def\eea{\end{eqnarray}}

\def\keV{\,{\rm keV}}
\def\MeV{\,{\rm MeV}}
\def\sec{\,{\rm sec}}

\def\cmm2{{\,\rm cm^{-2}}}
\def\cm2{{\,{\rm cm}^2}}
\def\cmm3{{\,{\rm cm}^{-3}}}
\def\gcmm3{{\,{\rm g\,cm^{-3}}}}
\def\kms{\,{\rm km\,s^{-1}}}

\def\he{$^4$He}
\def\li{$^7$Li}
\def\ber{$^7$Be}
\def\het{$^3$He}
\def\tri{$^3$H}
\def\yp{$Y_P$}

\def\fun#1#2{\lower3.6pt\vbox{\baselineskip0pt\lineskip.9pt
  \ialign{$\mathsurround=0pt#1\hfil##\hfil$\crcr#2\crcr\sim\crcr}}}

\def\kms{{\rm~km~s^{-1}}}

\def\ie{{i.e., }}
\def\etal{{\it et al.}}

\def\p3m{P$^3$M}

\def\la{\mathrel{\mathpalette\fun <}}
\def\ga{\mathrel{\mathpalette\fun >}}
\def\fun#1#2{\lower3.6pt\vbox{\baselineskip0pt\lineskip.9pt
  \ialign{$\mathsurround=0pt#1\hfil##\hfil$\crcr#2\crcr\sim\crcr}}}

\newcommand{\BBN}{{\rm BBN}}
\newcommand{\tbbn}{T_{\rm BBN}}
\newcommand{\hbbn}{H_{\rm BBN}}

\newcommand{\taun}{\tau_{\rm n}}
\newcommand{\dtbbn}{\Delta t_{\rm BBN}}

\newcommand{\hone}{H_{1}}

\begin{document}
\baselineskip 18pt

\title{Testing The Friedmann Equation: \\
The Expansion of the Universe \\
During Big-Bang Nucleosynthesis}
\author{
Sean M. Carroll$^{1}$ and Manoj Kaplinghat$^{2}$
\vspace{.2cm}\\
$^1$ Department of Physics and Enrico Fermi Institute\\
The University of Chicago, Chicago, Il 60637, USA\\
{\tt carroll@theory.uchicago.edu}\\
$^2$ Department of Astronomy and Astrophysics\\
The University of Chicago, Chicago, Il 60637, USA\\
{\tt manoj@oddjob.uchicago.edu}}

\maketitle

\begin{abstract}
In conventional general relativity, the expansion rate $H$ of a 
Robertson-Walker universe is related to the energy density by
the Friedmann equation.  Aside from the present day, the only
epoch at which we can constrain the expansion history in a 
model-independent way is during Big-Bang Nucleosynthesis (BBN).
We consider a simple two-parameter characterization of the 
behavior of $H$ during BBN and derive constraints on this 
parameter space, finding that the allowed region of parameter 
space is essentially one-dimensional. We also study the effects 
of a large neutrino asymmetry within this framework. Our results 
provide a simple way to compare an alternative cosmology to the 
observational requirement of matching the primordial abundances 
of the light elements. 
\end{abstract}


\section{Introduction}

Modern cosmology boasts a best-fit model which is in good agreement 
with a variety of data.  This model features a homogeneous and
isotropic spatially-flat universe comprised of approximately 
$5\%$~baryons, $25\%$~cold dark matter, and $70\%$~vacuum energy,
with a radiation energy density of about $10^{-4}$ at a temperature
$2.7^\circ$~K.  Along with a nearly scale-free spectrum of 
adiabatic density perturbations, these ingredients provide an
accurate match to observations of the recent expansion history 
\cite{snIa}, large-scale structure \cite{lss01}, 
the cosmic microwave background (CMB) \cite{cmb01}, and the
abundances of light elements as predicted by Big-Bang Nucleosynthesis
(BBN) \cite{burles01}.

As successful as the best-fit model has been, it falls short of
achieving an aura of inevitability due to certain problems of naturalness.
Foremost among these are the two cosmological constant problems:
why the vacuum energy is so small in comparison to its expected
value, and why its precise magnitude is so close to that of the
matter energy density today \cite{carroll}.  
In addition to these unsolved problems, there are problems for which
a popular solution exists but is lacking a firm experimental
footing: the flatness and horizon problems, which may be solved by
inflation, and of course the nature of the dark-matter itself.
Finally, amidst the large successes there are small failures of
the best-fit model in fitting the data, especially in the detailed
matching of structure on small scales to the predictions of
cold dark matter \cite{lsb}.  Meanwhile, numerous alternatives
to conventional cosmology have been studied.  To take one recent
example, brane-world models with large extra dimensions can give
rise to a variety of departures from four-dimensional
general relativity, with important consequences for the early
universe \cite{branes,chung}.

Given this situation, it is worth examining as carefully as possible
the theoretical assumptions of the standard cosmological model,
if for no other reason than to reassure ourselves that its apparent
fine-tunings are not pointing toward a radically different underlying
framework.  In this paper we consider the empirical evidence relating
to the Friedmann equation, the dynamical relation in general relativity
between the expansion rate of the Universe and the energy density in it.

The fundamental quantity in general relativity is the space-time
metric. We have some empirical evidence for the form of the metric
from astronomical observations.
We know that the Universe is expanding \cite{hubble}, \ie the
physical volume of the Universe is increasing with time. We also know
that the cosmic microwave background is isotropic to a part in $10^5$
\cite{cobe}. Therefore, it seems reasonable to assume that the spatial 
metric is isotropic. If we further assume that all observers (matter) in 
this expanding isotropic Universe see the same CMB isotropy, then it
follows \cite{egs} that the Universe is homogeneous (on scales much
larger than that of the largest collapsed structure). The metric is then 
uniquely determined. The form of this metric, called the
Robertson-Walker (RW) metric, is    
\be
  ds^2 = -dt^2 + a^2(t)\left[{dr^2 \over 1-kr^2} + r^2 d\Omega^2\right]\ ,
  \label{rwmetric}
\ee
where $a(t)$ is the scale factor and $k \in
\{-1,0,+1\}$ is the curvature parameter.  The Friedmann equation
(derived from general relativity for the RW metric) is
\be
  H^2 = \left({\dot a \over a}\right)^2 = {8\pi G \over 3}\rho 
  -{k \over a^2}\ .
  \label{feq}
\ee
$H = \dot a/a$ is the Hubble expansion rate 
and $\rho$ the energy density of the universe.  
In fact (\ref{feq}) is only the $00$ component
of Einstein's equations applied to (\ref{rwmetric}); we also have
another independent gravitational equation,
\be
  {\ddot a \over a} = -{4\pi G\over 3}(\rho + 3p) \label{feq2} \ ,
\ee
as well as an equation of energy-momentum conservation,
\be
  \dot\rho = -3 (\rho + p){\dot a \over a}\ ,\,
   \label{emcons}
\ee
where $p$ is the pressure.
In seeking to test this framework, we might hope to seek consistency
relations among these equations, which could be compared with data.
Unfortunately, it is always possible to find an energy density
$\rho(t)$ and pressure $p(t)$ which satisfy (\ref{feq}-\ref{emcons}).
Specifically, any given expansion history $a(t)$
and curvature $k$ is compatible with these equations if we choose
\bea
  \rho &=& {3\over 8 \pi G}\left[\left({\dot a \over a}\right)^2
  + {k \over a^2}\right] \\
  p &=& - {1\over 8\pi G} \left[2{\ddot a \over a} + \left({\dot a
  \over a}\right)^2 + {k\over a^2}\right]\ .
\eea
The crucial point here is that a perfectly smooth component of
energy and pressure cannot be detected in any way except for its
influence on the expansion rate, so we have no independent constraint
on such a source.  While invoking a perfectly smooth component to
fit an arbitrary behavior of the scale factor might seem like
cheating, this is exactly what we must do to reconcile
the strong evidence in favor of spatial flatness with the
similarly strong evidence that the amount of clustered matter falls
far short of the critical density in the current universe.  In this
sense, the set of equations (\ref{feq}-\ref{emcons}) are,
strictly speaking, untestable, without some prior expectation for
the nature of $\rho$ and $p$.

Instead, we can characterize what observations can tell
us about the behavior of the scale factor in a model-independent
way, so that any specific alternative theory can be straightforwardly
compared with the data.  There are only two eras of the universe's
history in which such a characterization is possible: the recent
universe, and the BBN era.  In the recent universe,
information about the behavior of the scale factor can be obtained
in a variety of ways, most directly by comparing the distance
of faraway objects to their redshifts, as is done in the 
supernova studies that first revealed the acceleration of the
universe \cite{snIa}.  During BBN, the expansion rate directly affects
the relative abundances of light elements produced (as reviewed
below), so that the primordial abundances are a powerful constraint
on alternative expansion histories. 

Other important observable phenomena which are affected by the
expansion rate do not offer such a direct test, since they typically
involve the local behavior of gravity (evolution of perturbations)
in addition to its cosmological behavior (evolution of the scale
factor).  Examples include the growth of large-scale structure and
the imprinting of temperature anisotropies on the CMB.  
A theory which predicts deviations from the Friedmann
equation could generically predict deviations from the predictions
of general relativity on other scales.  Consequently, it is 
difficult to use the growth of structure and CMB anisotropies as
model-independent tests of the Friedmann equation, although any
fully-specified alternative theory might be tightly constrained
by these phenomena. However, if one assumes that the local behavior
of gravity as predicted by general relativity is correct, then structure 
formation and CMB anisotropies are a powerful probe of the expansion
history (and hence dark energy evolution) after the last scattering
epoch \cite{tegmark01}. 

In this paper we will study what kinds of expansion histories in
the early universe are compatible with the BBN explanation of the light
element abundances.  Other works have constrained the
energy density \cite{kernan94} or value of Newton's constant
\cite{kolb86} during BBN, assuming the Friedmann equation, 
or have derived BBN constraints on specific scalar-tensor
theories \cite{santiago,damour} or have  
put limits on alternative cosmologies under the assumption that
the universe has undergone a consistent power-law evolution from
very early times \cite{kap00}. 
An estimate of the constraints imposed by BBN (by considering the change
in the neutron-proton freeze-out temperature) and structure formation on
cosmologies where the energy density in the universe varies as some
power of $H$ has also been calculated \cite{chung}. 
Our approach will be not to assume any specific behavior of
the scale factor over long periods, but instead to introduce a
two-parameter family of evolution histories which we take to be valid 
only in the vicinity of BBN. We find that a variety of alternative 
cosmologies can be consistent with observations, although they
comprise essentially a one-dimensional region in our two-dimensional
space of possibilities.  

\section{Parameterizing the scale factor during BBN}

We will consider theories in which the field equations for
the metric may be different from those of general relativity,
but we assume that test particles will still follow geodesics
of this metric.  From this assumption it follows that the
energy of a relativistic particle redshifts as $1/a$.  The
photon temperature will be proportional to $1/a$ in the absence
of entropy creation; however, $e^\pm$ annihilation can act
as a significant entropy source.  We therefore use the neutrino
temperature $T$ as a measure of the scale factor, since $T
\propto 1/a$ to excellent accuracy during the epoch under
consideration.  Of course these statements rely on our
decision to only consider changes in the gravitational 
dynamics, not any particle-physics processes.

The process of conventional BBN spans temperatures
between the freeze-out of weak interactions, $T_f = 1\MeV$,
and the synthesis of helium, $T_{\rm BBN} = 60\keV$;
this corresponds to a change in the
scale factor by slightly more than one order of
magnitude.  It will therefore be reasonable to approximate the 
expansion rate during this interval, so long as it is not wildly 
oscillating or somehow finely tuned, as a simple power law:
\be
  H(T) = \left({T\over 1 \MeV}\right)^{\alpha} \hone\ .
  \label{param}
\ee
Any expansion history is parameterized by the coefficient $\hone$ and
the exponent $\alpha$. (One can view this as a Taylor expansion in
$\log(H)$, to first order in $\log(T)$.) We do not extend this
parameterization beyond the BBN epoch, nor are we proposing any models
in which (\ref{param}) is predicted (but note that all the models in
\cite{branes,chung} are well described by this expansion law); this is a
phenomenological study of the expansion rate during BBN. It should also
be noted that there might exist models which are not accurately
described by power-law expansion during the BBN era \cite{damour}; in
such cases it would be necessary to examine each model individually.  

In standard BBN weak interactions freeze out close to $T_f = 1\MeV$,
at which point $n/p(T_f)\approx 1/6$. At temperatures lower than $T_f$,
the only change in the neutron to proton ratio occurs due to free decay
of neutrons with a lifetime of $\taun = 887 \sec$. At temperatures much
lower than freeze-out,
it becomes possible to synthesize helium in appreciable
quantities. This happens at a temperature of $\tbbn = 60\keV$, and
essentially all the remaining neutrons (one per seven protons) are bound
into helium nuclei. It should be noted that this temperature ($\tbbn$)
is much lower than the binding energies of all the light
elements. Clearly, the large photon to baryon ratio (which implies a
large photo-dissociating background) is partly responsible for this
\cite{kt}. The rate at which  $n\rightarrow$~\he\ conversion occurs
depends on the destruction rate of deuterium which in turn depends on
the binding energy of deuterium, the formation rate of tritium and the
baryon to photon ratio. In standard BBN, it turns out that the
conditions for helium synthesis are just right when the helium formation
rate is larger than $1/\taun$ \cite{esmail91}. This picture breaks down
in the unconventional expansion histories we consider.     

An understanding of how alternative models affect light-element
abundances can be expressed in terms of the behavior of three
characteristic quantities: the freeze-out temperature $T_f$ (which sets
the initial neutron/proton ratio), the time interval between 
freeze-out and nucleosynthesis, $\dtbbn = t_\BBN - t_f$ (during
which neutrons decay), and the Hubble parameter at nucleosynthesis,
$\hbbn$ (which affects the efficiency with which neutrons are
converted to helium, and thus determines the deuterium and lithium
abundances). 
Note that $\dtbbn \approx t_\BBN$, since $t_f << t_\BBN$; this will
continue to be true in the non-standard models we study. Assuming only
that $H$ is monotonically decreasing even prior to the epoch when this
formula is relevant, we will have 
\be
  t_\BBN = {1 \over \alpha \hbbn}\ .
\ee

The predictions of Big-Bang nucleosynthesis are given as ratios of
abundances of the light elements -- neutrons (n), protons (p or H), 
deuterium (D), \tri, \het, \he\ and \li. One quotes the ratio of the
number density of the light elements to the number of protons, except in
the case of helium where the accepted practice is to quote its mass
fraction (\yp). The absolute density of these light elements is
set by the absolute density of baryons. The only input parameter in
standard BBN is the baryon density quoted as its ratio ($\eta$) to
the number density of photons. It is useful to define
$\eta_{10}=10^{10}\eta$ since the number density of baryons (compared to 
that of photons) is very small. Except for a brief period when electrons
and positrons annihilate, $\eta$ remains constant (provided no other
entropy changing interactions come into equilibrium). 

We will adopt observationally allowed ranges of light elements ratios in 
keeping with the inferences (but slightly on the conservative side) of
Olive \etal\ \cite{olive00}. The ranges are $0.228 \leq Y_P \leq
0.248$, $2 \leq 10^{5} \times D/H \leq 5$, and $1 \leq
10^{10}\times{}^7Li/H \leq 3$. The low deuterium value has been adopted
since it seems to be favored by current data \cite{kirkman01}. 

The baryon density is not as well constrained observationally as the
ratios of light element abundances. One reason for this is that we can
only inventory the baryons that we see and most of the baryons could be
dark. Even putting a lower bound on the baryon density is a difficult
task. Persic and Salucci \cite{persic92} estimated the density of
baryons in the Universe (in terms of $\eta_{10}$) to be $\eta_{10}\geq
0.3$. In converting the baryon density to $\eta$ we have assumed that
$\eta$ has remained constant since BBN (as it does in the standard
model), and that $H_0\geq 60\kms$ \cite{freedman01} where $H_0$ is the
present expansion rate of the Universe. Fukugita \etal\ \cite{fuku98}
also did a baryon inventory and their bound on the baryon density
translates to $\eta_{10} \geq 1.8$, a much higher lower bound on the
baryon density. We will use both these lower bounds as a way of showing
how the allowed region changes with the bound on the baryon density.  

Without recourse to a specific theory of gravity it is not possible to
put an upper bound on the baryon density. For our purposes, we will
assume the following ranges for the baryon density -- a tight bound
corresponding to $1\leq \eta_{10} \leq 10$ and a broader range with
$0.5 \leq \eta_{10} \leq 50$. We will also comment on what happens if
the upper bound on the baryon density is increased.

\section{Helium constraints}

The requirement that about 24\% helium by mass be synthesized imposes
severe constraints on the expansion history for temperatures around
$\tbbn$. To understand these constraints, it proves useful to look at the 
effects of changes in the parameters $\hone$ and $\alpha$ 
of (\ref{param}) in the vicinity of the
standard model. We first concentrate on the effect of changing $\hone$
at fixed $\alpha$. As mentioned, the important considerations are the
freeze-out temperature $T_f$, and the time interval between freeze-out
and nucleosynthesis $\dtbbn \approx t_\BBN$.  (In the vicinity of
standard BBN, small changes in the expansion rate $\hbbn$ will not
be important, since essentially all neutrons are converted to 
helium.)  Increasing $\hone$ increases
the expansion rate at every temperature; thus, freeze-out will occur
earlier ($T_f$ will be higher) leading to a larger initial
neutron/proton ratio. At the same time, the faster expansion rate
leads to a decrease in $t_\BBN$, leaving less time for neutrons to
decay. Both these effects go  in the same direction and hence the
neutron/proton ratio at $\tbbn$ increases with increasing $\hone$,
leading to a higher helium abundance.  

One can analyze the effect of changing $\alpha$ at fixed $\hone$ in a
similar manner. Again, the relevant considerations in the vicinity 
of standard BBN are the behaviors of $T_f$ and $\dtbbn$.
Since $\hone$ corresponds to a temperature close to 
freeze-out and before helium synthesis, increasing $\alpha$
means that the expansion rate will be lower during
nucleosynthesis. In particular this implies that $\hbbn$ will be lower 
(\ie helium synthesis happens later), and hence that the time interval
$\dtbbn\propto 1/\hbbn$ will be longer.  Consequently, increasing
$\alpha$ at fixed $\hone$ gives neutrons more time to decay between
$T_f$ and $T_\BBN$, working to decrease the helium abundance for values
close to the standard ones. 

\begin{figure}[htb]
  \plottwo{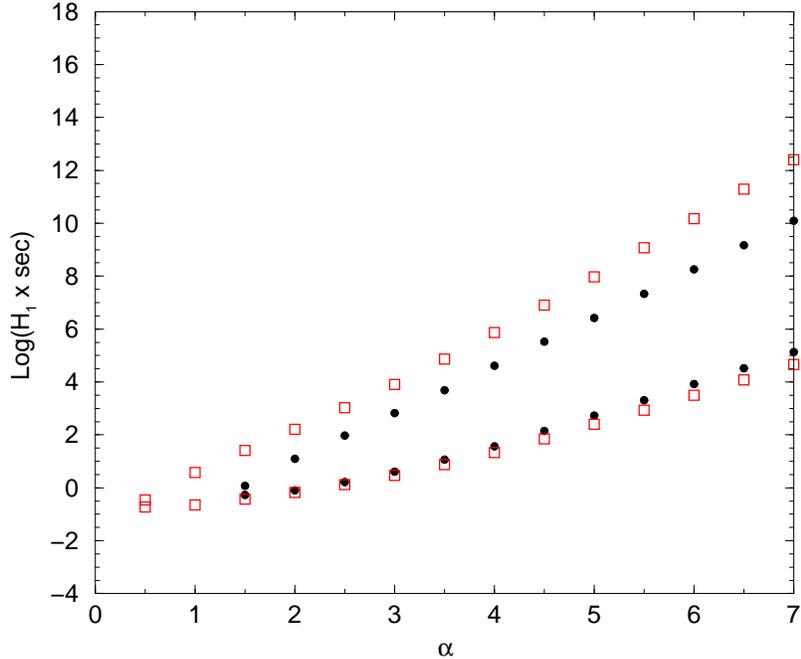}{.65}
  \caption{Contours representing constant helium in the 
  $\alpha$-$\hone$ plane, for $Y_P = 0.24$.  We have chosen
  $\eta=10^{-10}$ (filled circles) and $\eta=10^{-9}$ (squares) for
  purposes of illustration. }
  \label{helium}
\end{figure}

We can therefore balance the effects of increasing $\hone$
against those of increasing $\alpha$, to obtain a constant
helium abundance for small deviations from the standard
picture.  In Figure \ref{helium} we plot contours of constant
\he\ in the $\alpha$-$\hone$ plane.
This figure reveals the expected behavior in the vicinity
of the standard model, but we see that the curves turn over,
and at fixed $\alpha$ there is a larger value of $\hone$ for
which the helium abundance is the same.
This feature has been previously noted in the context of
exploring the effects of very large numbers of light neutrino
species (and thus $\hone$) \cite{ossty} and in the explorations of
BBN constraints on scalar-tensor theories of gravity \cite{peebles}.
For very large $\hone$, the freeze-out temperature is sufficiently
high that $n/p(T_f)$ asymptotes to unity; $\dtbbn$ continues
to decrease as $\hone$ increases, leading to an ever-larger
neutron fraction at nucleosynthesis.  However, $\hbbn$ also
increases and eventually the expansion rate at nucleosynthesis
is so large that there is not enough time to efficiently
turn the neutrons into helium.  Thus, given the success of
standard BBN, there will be an additional larger value of
$\hone$ (at the standard value $\alpha=2$) for which the
helium abundance is the same. This argument can be generalized not just
to other values of $\alpha$ but also to the other light element
abundances. One expects (at fixed $\alpha$) to get two values of $\hbbn$ 
which produce the same abundance.

\section{Deuterium and Lithium constraints}

\begin{figure}[htb]
  \plottwo{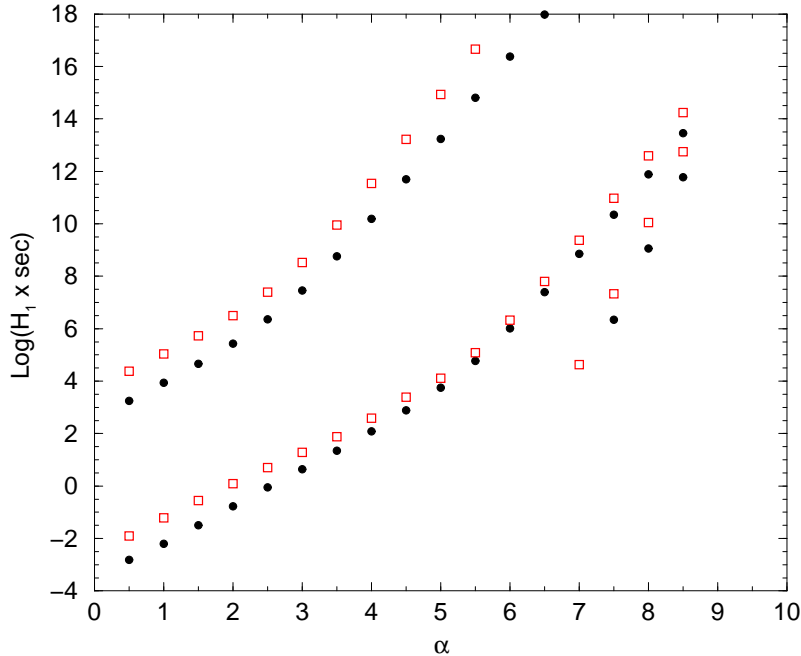}{.65}
  \caption{Contours of constant deuterium, D/H$=3\times 10^{-5}$, at
   $\eta=10^{-10}$ (filled circles) and $\eta=10^{-9}$ (squares).}  
  \label{deuterium}
\end{figure}

The deuterium abundance provides an important additional constraint
on the allowed parameter space. In the standard picture deuterium is a
by-product of neutrons getting burnt to helium. Since the abundance of
deuterium is set by its destruction rate and the time available for this 
destruction, it is extremely sensitive to the expansion rate at
temperatures close to $\tbbn$.
Around the standard model, D/H increases with $\hone$ for
a fixed $\alpha$, since the time available for nucleosynthesis
decreases. If one fixes $\hone$, the decrease in post-BBN ($T \la
\tbbn$) expansion rate causes D/H to decrease with increasing
$\alpha$. Once again we see that there is a trade-off to be made between 
$\alpha$ and $\hone$, and hence it is possible to increase $\alpha$ and
$\hone$ simultaneously such that we end up with the same deuterium
abundance. This is borne out by the contours of constant
deuterium plotted in Figure \ref{deuterium}. Note that one expects
exactly the opposite behavior with respect to changes in $\alpha$ (with
$\hone$ fixed), and $\hone$ (with $\alpha$ fixed) for the upper branch of
contour plot. Essentially, the abundances in the upper branch of the
contour plot increase if the time available for nucleosynthesis goes up.

{}From Figure \ref{deuterium} it is clear that the lower branch of the
contour plot turns over at large $\alpha$. This can be traced to 
the appearance of a new production channel.
As $\alpha$ increases without a corresponding
exponential increase in $\hone$, the time available for nucleosynthesis
becomes very large, and consequently deuterium is almost completely
destroyed. In the lower branch this implies that beyond a certain
value of $\alpha$ it would not be possible to produce a reasonable D/H
abundance. However, if the age of the Universe gets large enough it
becomes possible to produce deuterium by pp fusion, in a manner
similar to what happens in the sun. This pp burning explains the turn over
towards smaller values of $\hone$ in the lower branch of the deuterium
contour. For this solution to work, the age during nucleosynthesis has
to be of the order of billions of years. We, therefore, consider this
solution highly unlikely and do not include it in our analysis.

For our parameter space around the standard model we expect the same
behavior with respect to $\eta$ as in standard BBN. All other things
being the same, increasing $\eta$ decreases deuterium abundance while
increasing $Y_P$ due to more efficient nucleosynthesis. Thus at fixed
$\alpha$ if $\eta$ is increased, then one needs to decrease $\hone$ to
lower the helium abundance back to where it was. In the case of
deuterium, if $\eta$ is increased at fixed $\alpha$, then one needs to
increase $\hone$ in order to increase D/H back to where it was. 

Note that for the upper branch one would have to increase $\hone$ at
fixed $\alpha$ to compensate for an increase in $\eta$ for {\em both}
helium and deuterium, since both $Y_P$ and D/H increase with $\eta$ in
the upper branch.

\begin{figure}[htb]
\plottwo{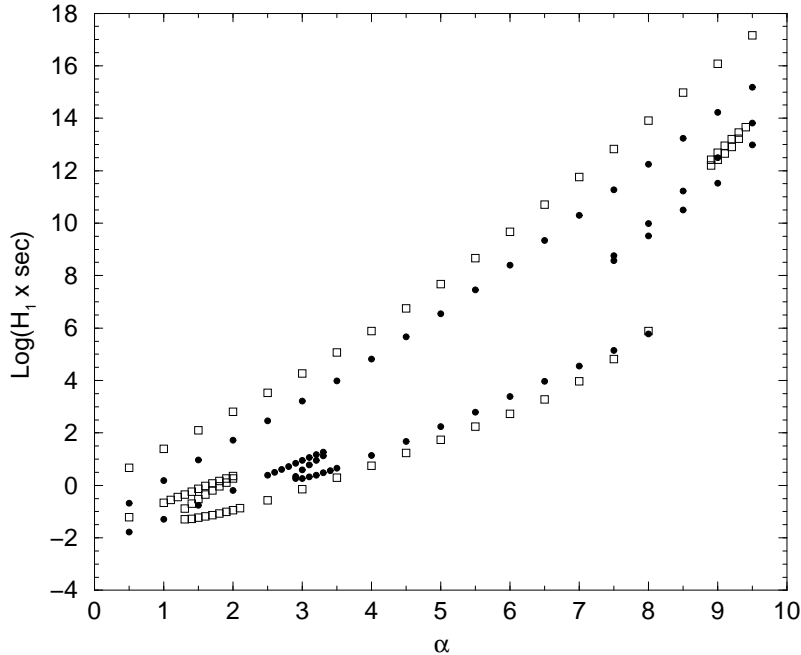}{.65}
  \caption{Contours of constant lithium, \li/H$=3\times 10^{-10}$, at
   $\eta=10^{-10}$ (filled circles) and $\eta=10^{-9}$ (squares).} 
  \label{lithium}
\end{figure}

We can infer the primordial abundance of \li\ very accurately from
poor-metallicity stars \cite{olive00}. Therefore, it provides an
additional constraint on our parameter space. For most of the
parameter space, getting helium right automatically ensures that
lithium is near the observed abundance. Even so, with the very tight
bounds on \li/H that we impose, lithium does provide strong constraints. 
Figure \ref{lithium} shows contours of
constant lithium in the $\hone$--$\alpha$ plane. Its behavior with
respect to a change in $\eta$ follows that of helium. 

The \li\ iso-abundance contour has two interesting features not
present in the helium contour, which can be traced to the fact that \li\
has two distinct production channels. \li\ can be produced through the 
reaction  \he(\tri,$\gamma$)\li, and through \he(\het,$\gamma$)\ber\
with the subsequent beta-decay of \ber. Fixing our attention on the
lower branch of the lithium contour, as we move from low to high
$\alpha$, the dominant production channel changes from the direct \li\
channel to the indirect \ber\ one. The kink in the lower branch of the
contour is at values of $\alpha$ where this transition takes place. At
large $\alpha$ and $\hone$, there is a small closed contour below the
upper branch. For a fixed $\alpha$ in the region spanning the width of
the small closed contour, there are three values of $\hone$ which
produce the required lithium abundance. The solution with the lowest
value of $\hone$ produces \li\ through the indirect \ber\ channel. In
the upper branch of the contour \li\ production is through the direct
channel, for all values of $\alpha$.  A larger $\hone$ means
shorter time for nucleosynthesis and since \tri\ is easier to burn than
\het, the direct channel dominates.  

\section{Combined constraints}

\begin{figure}[htb]
  \plottwo{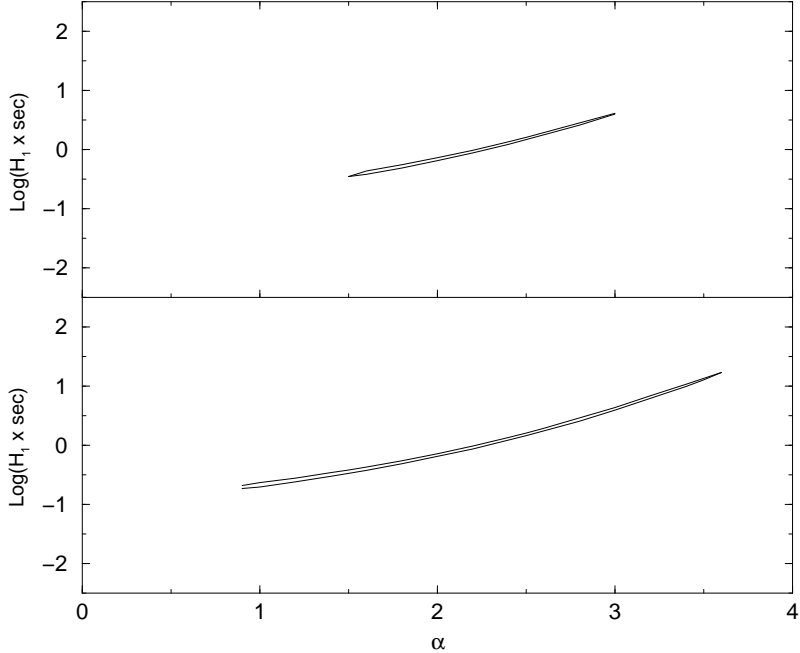}{.65}
  \caption{Allowed values of $\alpha$ and $\hone$ which lead to
   acceptable light element abundances. Upper panel shows the 
   allowed parameter space when we constrain $\eta_{10}$ to be between 1 
   and 10. The lower panel shows the allowed parameter space for
   $0.5 \leq \eta_{10} \leq 50$.}
  \label{final}
\end{figure}

Putting together all of our constraints yields the allowed
region shown in Figure \ref{final}.
We note that the allowed region is essentially
one dimensional, and is characterized
by an almost linear relationship between $\log(\hone)$
and $\alpha$. This tells us that there must exist a temperature
$T_c$ at which the expansion rate $H_c$ is approximately fixed for
all the models in the allowed range. In other words, we must have
\be
\hone = H_c\,\left(T_c/\MeV\right)^\alpha; 
\quad H_c=(0.039 \pm 0.013) /\mathrm{sec} 
\quad \mathrm{at} \quad 
T_c=0.2\,\MeV\,.
\label{Hc}
\ee
The range of $H_c$ quoted in Eq. \ref{Hc} corresponds to our
conservative limits on the baryon to photon ratio. For the case with
the tighter bound on baryon density, we obtain $H_c=(0.03 \pm 0.004)
/\mathrm{sec}$.
The value of $T_c$ reflects of the physics involved. \he\ is
primarily sensitive to $\hone$ (as well as to $\alpha$), 
while the other elements (which can be
viewed as by-products of the neutron to helium burning process) are
mainly sensitive to $\hbbn$. Thus, not surprisingly, we have 
$1\MeV > T_c > \tbbn$. 

The coincidence that $T_c$ is so close to the $e^\pm$ annihilation
temperature ($\simeq m_e/3 = 0.17 \MeV$) implies that the expansion rate 
of the Universe during $e^\pm$ annihilation must have been close to the
standard BBN value. This in turn implies that the cosmic microwave
background will retain a black body spectrum through the annihilation
epoch (as in standard BBN) in all the viable non-standard
expansion histories we have identified. 
  
The effect of a change in the assumed range of the baryon density on the 
bounds on $\alpha$ is visible in Figure \ref{final}. Decreasing the
lower bound on $\eta$ increases the upper bound on $\alpha$, while
allowing for larger baryon densities implies a smaller lower bound on
$\alpha$. The reason behind such behavior is not hard to 
understand.  Given
that the expansion rate at 0.2 $\MeV$ is approximately fixed, a viable
expansion history with larger $\alpha$ has lower $\hbbn$. A lower
$\hbbn$ (\ie more time for nucleosynthesis) at fixed baryon density
implies a lower deuterium abundance. Thus one requires a smaller
baryon density to get deuterium abundance right in the limit of large
$\alpha$. Similarly, decreasing
the lower bound on $\alpha$ requires larger baryon densities.

As discussed before, we do not really have an observational upper bound
on the baryon density of the Universe in the absence of a theory of
gravity. The lower bound is on a more sound footing because we can count
the visible baryons.
Thus it seems pertinent to ask what happens if one allows for larger
cosmological baryon densities. From our arguments above we would expect
viable models with smaller values of $\alpha$. However, increasing
$\eta_{10}$ above 50 (panel 2 of Figure \ref{final}) has {\em no
effect} on the allowed range of $\alpha$. The reason behind this result 
is best understood through the following qualitative picture. At smaller
values of $\alpha$, and therefore larger $\hbbn$, one has a higher
deuterium abundance if the baryon density is fixed. To lower the
deuterium abundance to required levels one has to increase the baryon
density. However, a larger baryon density (keeping $\alpha$ fixed) will
lead to a larger helium fraction. Hence below a certain value of
$\alpha$ (which depends on the constraints on the abundances) it
is not possible to simultaneously obtain the correct amounts of
deuterium and helium. This is also the reason why the allowed contour in
the lower panel of Figure \ref{final} does not appear to close. 

\section{Neutrino asymmetry}

One of parameters in cosmology on which we have little handle is the
neutrino asymmetry, i.e., the excess of neutrinos over anti-neutrinos or 
vice versa. This excess or deficit can be quantified by the neutrino
chemical potential, $\mu$, which enters into the Fermi-Dirac
distribution function, given by $1/[1+\exp(p/T+\mu/T)]$. The chemical
potential of anti-neutrinos is $-\mu$. The ratio $\mu/T$ is an
invariant, as long as there are no entropy-changing processes which
involve neutrinos. 

There are no compelling experimental or theoretical reasons to expect
the neutrino (lepton) asymmetry ($\propto (\pi^2+\mu^2/T^2)\mu/T$) to be
orders of magnitude different from the baryon asymmetry. In fact, one
might very well argue that it would only be natural for the baryon and
lepton asymmetries to be comparable if they were formed by the same
processes.  
However, we have no concrete experimental hints to back up our
theoretical prejudice. Moreover, in our framework, the value of $\mu/T$
is virtually unconstrained (but for BBN) since the standard cosmological
bounds \cite{freese83} do not apply. Note that in order to probe  
$\mu/T \sim 1$ regime, one would require experiments which are capable
of measuring neutrino energies down to $10^{-3}$ eV. As we shall shortly
see, BBN even without the assumption of standard gravity provides us
with constraints on the value of the electron neutrino chemical
potential in the $|\mu/T|={\cal O}(1)$ regime. 

Changing the neutrino chemical potential affects the neutrino number
density. In standard BBN, a change in the neutrino number density
affects the expansion rate and hence the abundances. Clearly, this
effect is non-existent in our study. The electron neutrino chemical
potential also affects the $n \leftrightarrow p$ reaction rates. This is 
the effect we shall be concerned with in this section. From here on,
$\mu/T$ will implicitly refer to the electron neutrino chemical
potential. 

\begin{figure}[htb]
  \plottwo{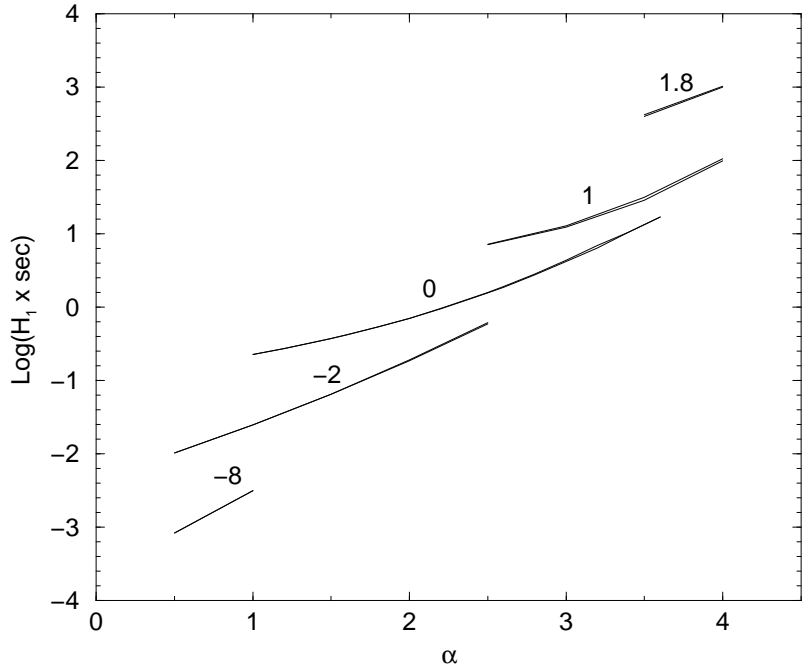}{.65}
  \caption{Allowed values of $\alpha$ and $\hone$ for $Y_P=0.244$ and
        $0.5 \leq \eta_{10} \leq 50$. The contours are labelled by the
        value of $\mu/T$. No solutions consistent with our constraints 
        were obtained for $\mu/T > 2$ and $\mu/T < -10$.} 
  \label{phinue}
\end{figure}

Detailed calculations of the effect of non-negligible $\mu/T$ (i.e.,
$|\mu/T| \ga 1$) on BBN have been performed \cite{steigman92}. The
salient points are (1) the equilibrium value of n/p is modified by a
factor of $\exp(-\mu/T)$ from that when $\mu=0$, (2) the neutrino
decoupling and n-p freeze-out temperatures are changed. The first point
clearly implies the constraint, $\mu/T < 2$, 
if a helium-4 mass fraction of 24\% 
is to be synthesized. In Figure \ref{phinue} we plot the
allowed contours in $\alpha$ -- $\hone$ plane. The contours are
labelled by the value of $\mu/T$. We have restricted $Y_P$ to being
24.4\%. 
Allowing for our previously adopted range in $Y_P$ will not change the
result much as is apparent from the fact that the allowed contours in
Figure \ref{final} are practically 1-dimensional.  
We note that if $\alpha$ is set equal to 2, then we are back to standard
cosmology; Figure \ref{phinue} then provides us with the constraint on
the number of relativistic degrees of freedom ($\propto \hone$)
\cite{steigman92}. 

The principal feature in Figure \ref{phinue} is the trend of the
required value of $\hone$ being larger for larger $\mu/T$. At larger
$\mu/T$, the equilibrium value of n/p is smaller, and hence to compensate 
for that $t_\BBN$ must be smaller or $\hone$ larger. For $\mu/T < -10$ no 
solutions were found, which owes to the fact that at the small values of 
$\hone$ required to produce the right helium-4, $t_\BBN$ gets so large
that one cannot synthesize acceptable amounts of $D/H$. 

\section{Discussion}

\begin{figure}[htb]
  \plottwo{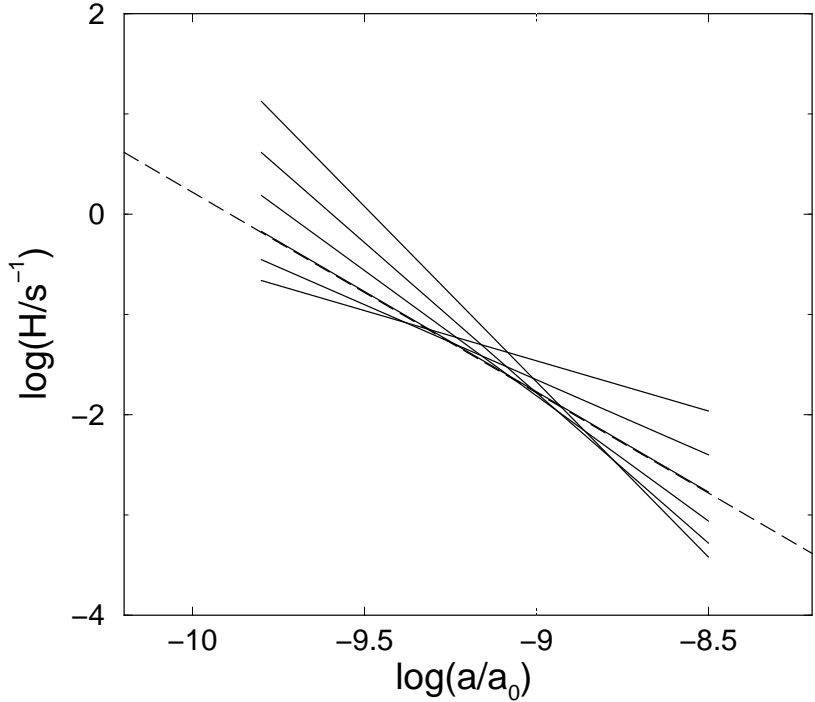}{.65}
  \caption{The expansion rate $H$ as a function of scale factor
  $a$, for the family of allowed models in our parameterization
  (corresponding to our conservative limits on the baryon/photon
  ratio and negligible electron neutrino asymmetry).
  The dashed curve represents the standard model, and the solid curves
  are different allowed histories.  It is clear that there is a
  value of the scale factor $a_c/a_0 \simeq 8.5\times 10^{-10}$ at which 
  the allowed range in $H$ is minimized; this corresponds to a
  temperature $T_c \simeq 0.2 \MeV$.}
  \label{hvsa}
\end{figure}

We have shown that the viable (\ie satisfying BBN abundance
constraints) expansion histories can be well approximated by the range
of allowed values of $\alpha$, for a fixed value of $\mu/T$. 
Lets concentrate on the more ``natural'' $|\mu/T| << 1$ (equivalent to
$\mu=0$) case.  
Given the allowed range of $\alpha$ one can generate the allowed
range of expansion rates at any temperature from about an MeV to 50~keV
by using 
\be
H(T)=H_c\,\left(T/T_c\right)^\alpha\,.
\ee
We have plotted the history of these viable non-standard Universes in
Figure \ref{hvsa}. 

The most interesting result of our study is that there is a
range of expansion histories that are compatible with the
observed light-element abundances.  There is therefore room for
substantial deviation from the standard cosmological model at early
times, while remaining consistent with empirical evidence.
On the other hand, it is encouraging to note that our allowed
region (essentially one-dimensional) is only a small volume of the
entire parameter space. In this sense it would be unlikely to find
that any particular model
was both very different from the standard picture, and consistent with
the data.

We have also considered the effect of a large neutrino chemical
potential. Independent of general relativity, BBN constrains the value
of the electron neutrino chemical potential to $-10 < \mu/T < 2$. Of
course, this increases the allowed range of $\hone$ considerably, as can
be gauged from Figure \ref{phinue}.   

Throughout this discussion we have assumed tight but reasonable bounds
\cite{olive00} on the light element abundances. Our quantitative
results are sensitive to the assumed ranges. As observations improve
further, it will be possible to precisely pin down the expansion
history of the Universe during the epoch of nucleosynthesis.  

\section*{Acknowledgments}
We would like to thank Lloyd Knox and Michael Turner for useful
conversations.  This work is supported in
part by the Alfred P. Sloan Foundation, the David and Lucile
Packard Foundation, and the U.S. Department of Energy.

\end{document}